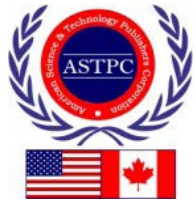

# CAN JSP CODE BE GENERATED USING XML TAGS?

Neha Bothra
Department of Computer Science and Engineering,
Institute of Engineering and Management, Kolkata, India
Email: nehabothra93@gmail.com

Kritika Jain
Department of Computer Science and Engineering,
Institute of Engineering and Management, Kolkata, India
Email: jkritika18@gmail.com

Sanjay Chakraborty
Asst. Professor, Department of Computer Science and Engineering,
Institute of Engineering and Management, Kolkata, India Email:
Sanjay.chakraborty@iemcal.com

*Abstract*— **Over the years, a variety of web services have started using server-side scripting to deliver results back to a client as a paid or free service; one such server - side scripting language is Java Server Pages(JSP). Also Extensible markup language (XML), is being adopted by most web developers as a tool to describe data .Therefore, we present a conversion method which uses predefined XML tags as input and generates the corresponding JSP code. However, the end users are required to have a basic experience with web pages. This conversion method aims to reduce the time and effort spent by the user (web developer) to get acquainted with JSP. The conversion process abstracts the user from the intricacies of JSP and enables him to focus on the business logic.**

## INTRODUCTION

Since the birth of dynamic web (with CGI), the web has developed tremendously and is powered by server side technologies like Java Server Pages (JSP), Active Server pages(ASP) ,PHP and others. But the reason why we have chosen JSP as the target code of our converter is because of the advantages that it offers over some of its competitors. Firstly, it uses Java, a language that already has a large skill set of developers and a massive amount of functionality available. Secondly, JSP is portable to other operating systems and Web servers , are platform independent. Also, JSP pages enable cleaner and more module application design because they provide a way to separate applications programming from web page design. This means personnel involved in web page design do not need to understand Java programming language syntax to do their jobs.

Though it provides a robust and scalable platform, JSP has a steep learning curve. Therefore, in this paper we present a code conversion process that takes an xml file as input and generates the corresponding JSP code. Our XML has been the choice of input because of it allows us to define our own tags as per requirements. The input xml file consists of input logic (whose JSP scriptlet needs to be generated) and the output is written out to a separate JSP file.

The well-formedness (W3C XML specifications) and validity of the input is checked against an XML Schema Definition (XSD) [5]. The validated input is then passed on to an operation handler .Our motivation for the converter has been derived from the conversion process described in "automatic conversion"[1].

For the conversion process we have utilized SAX (Simple API for XML) parser for both the validation and operation handling process. SAX has been chosen over DOM (Document Object Model) parser. The reason for choosing SAX is that for our converter sequential processing suffices and there is no need to retain all the information of the xml document. These reasons make SAX parser approach faster and space efficient in comparison to DOM [2].

The source-code for the converter has been developed using java programming language. In our converter we have incorporated basic error handling mechanism.

## PROPOSED WORK

JSP elements fall into four groups: directives, scripting, comments and actions. In our proposed conversion process, we mainly focus on the scripting and action tags.JSP scripting consists of the following:

- Declaration: used to declare instance, class variables and method used in a JSP page. Syntax: <%! declaration %>

- Scriptlets (inline java code): general business logic and code goes in here.
  Syntax : <% code %>

Action tags used are:

- forward :syntax:
- include: syntax:

We have used the self descriptive property of XML to define a specific set of tags. These tags are used to define the algorithm(or business logic) of the required JSP page. The input xml file must be well-formed meeting the requirements of having a closing tag for every element and must have a root element.

TABLE I.  Few of the predefined XML Tags

| Tags | Description |
|------|-------------|
| <array>data_type var_name[x] </array><br><var> var_name =  var_value </var> | x: non-zero size of array. Index goes from 0 to x-1. var_name : non-array variables. |
| <read>var_name <object>request/session </object> <type>parameter/attribute </type> <name>param_name </name></read> | To read external input like: a=request.getParameter("t l") from an html textbox named "tl" |
| <out> <write>text</write> <writev>var_name </writev> </out> | write: Displays text. writev: Displays variable. out: Displays all in a single line. |
| <dB> <driver>driver_name </driver> <url> db_url </url> <uid>user_id</uid> <pwd> password </pwd> <conn_name>var_name </conn_name> </dB> | Data connection details. To connect to different databases we use a seperate dB tag for each database connection. |
| <s> if (condition) then </s> . <s> endif </s> | A basic conditional statement. The condition must be enclosed within parenthesis. |
| <s> loop from i=a to (condition) step x </s> .. <s> endloop </s> | This is a basic for loop statement. The variable i must be declared before being used. |
| <function> <header> return_type function_name(real x,integer y, real a[], integer b[],....) </header> . </function> | This defines a function. A function starts with the function tag. It is then followed by the header- which comprises the function name, followed by the list of arguments. |
| <redirect>URL</redirect> | This tag redirects control to the page mentioned within the tag. |
| <ps> <query> statement_name=".." </query> <set> data_type (arg.no ,value)</set> <result> var_name  </result> <get>var_name= | Used to implement a database operation(like insertion) Text after the statement name will give the sql query type. Argument Number will be |

| data_type(argument_no) </get> </ps> | the column number of the table in database. |
|------|------|
| <class>class_name object_name <pname> param_name=param_value </pname> </class> | Example:<class>Date d</class> pname is optional |
| <include>file_name </include> | To include a file dynamically. |
| <forward>file_name <pname> Param_name=param_value </pname></forward> | pname is optional. |
| <session> <set> param_name=param_value</set> </session> | Setting a session attribute. |

Some of the predefined tags has been shown in the Table 1.There are some other tags which has not been shown in the table .The user is required to give his input within those tags.

To ensure that the XML input file is valid as per the table shown above, we use XSD or XML schema [5]. An XSD describes the structure of an XML document .An XSD has been used instead of DTD because it is extensible to additions ,it is easier to define restriction on elements and describe the content of document [3].

```
<xs:complexType name="out">
    <xs:choice minOccurs="0" maxOccurs="unbounded">
        <xs:element name="write" type="xs:string"/>
        <xs:element name="writev">
            <xs:simpleType>
                <xs:restriction base="xs:string">
                    <xs:pattern value="\s*_*[A-Za-z][\w_]*\s*"/>
                </xs:restriction>
            </xs:simpleType>
        </xs:element>
    </xs:choice>
</xs:complexType>
```

Fig. 1: Part of the designed XSD

In the XSD we have applied restrictions using regular expressions [4] on the input given within the tags and the order in which the tags can be nested as illustrated in Fig. 1.

Validation of input against the schema is performed using SAX (Simple API for XML) parser [6]. SAX parser is an event-based serial-access mechanism for accessing XML documents. A SAX parser reads an XML document as a stream and invokes call back functions provided by the application. Compared to a SAX parser, a DOM parser is much more complex, and hence, much slower [2].Therefore we have used a SAX validation parser to validate the xml file. Following the successful validation of the input xml file against the XSD, we use an operation handler which generates the appropriate JSP scriplet corresponding to the particular xml tag.

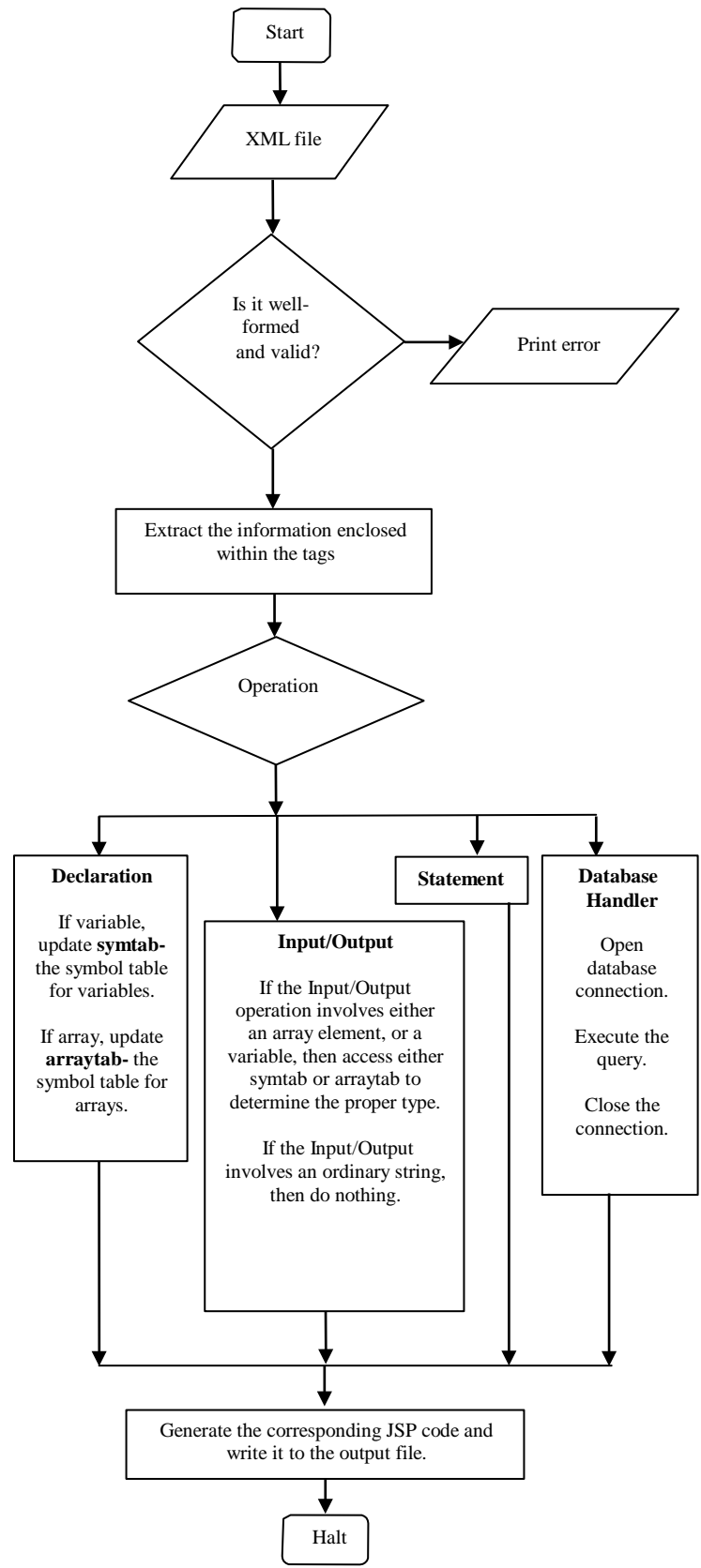

```
21      @Override
 8  ⊞   public void startDocument() {...7 lines }
29
30      @Override
 8  ⊞   public void endDocument(){...19 lines }
51      @Override
 8      public void startElement(String uri, String localName,
53          String qName, Attributes attributes) throws SAXException
54  ⊞      {...4 lines }
58
59      @Override
 8      public void characters(char ch[],
61  ⊞          int start, int length)throws SAXException {...26 lines }
81      @Override
 8      public void endElement(String uri, String localName,
            String qName)throws SAXException {...8 lines }
91
```

Fig 2: SAX Callback methods used in the operation handler

The operation handler itself has been designed using SAX parser to extract the start, end elements along with the text within the elements as shown in Fig 2.
Use of the SAX parsing callback methods to extract the information within the tags is adopted since there is no need to store the data or access it randomly .So it provides a faster and memory efficient approach.

The programming language used to develop the XSD validator and operation handler is java. The operation handler after identifying the type of tag takes the necessary action. Our converter has been designed to detect simple errors like repeated declaration or use of an undeclared variable, unclosed xml tags or improper input.
A separate handler has been designed for each category of tags .The generated code snippets are written out to a separate JSP file. The generated JSP file can be compiled separately. No changes are made to the original xml file.

```
<var> a ="" </var>
<read> a
<object>request</object>
<type>Parameter</type>
<name>t1</name>
</read>
<writev> a </writev>
```

Fig 3: Sample XML Input

```
String a="";
a=request.getParameter("t1");
System.out.println(a+"");
```

Fig. 4: Sample JSP Output

Fig 5: Flow-Chart representing operation Handler

The flowchart shown in Fig. 5 shows the categories in which the tags have been segregated in terms of the operation handling procedures.

## EXPLANATION WITH EXAMPLE

We now explain our work with an example.

```
1.  <?xml version="1.0" encoding="UTF-8"?>
2.  <root>
3.  <declare>
4.  <var> a="this is how!"  </var>
5.  </declare>
6.  <s> loop from xx = 2 to   10  step 2</s>
7.  <out>
8.  <write> the value is :</write>
9.  <writev> xx </writev>
10. </out>
11. <s>  endloop </s>
12. <dB>
13. <driver>com.mysql.jdbc.Driver</driver>
14. <url>jdbc:mysql://localhost/demo1</url>
15. <conn_name> conn </conn_name>
16. <uid>root</uid>
17. <pwd> password123 </pwd>
18. <excep_msg> oops...error !!!</excep_msg>
19. </dB>
20. <ps>
21. <var> b= 0</var>
22. <query> query="Update emp set phone=?
    and sal=? where eid=1011"</query>
23. <read> b
24. <object>request</object>
25. <type>parameter</type>
26. <name> t1 </name>
27. </read>
28. <set> int(1,b)</set>
29. <set> double(2,20000) </set>
30. <result> r </result>
31. </ps>
32. <s> if( r!=0) </s>
33. <write> Update successfull </write>
34. <s> else </s>
35. <write> Update unsuccessful</write>
36. <s> endif </s>
37. </root>
```

Fig 6: Input XML file: sample.xml

```
run:
Input validated successfully
BUILD SUCCESSFUL (total time: 0 seconds)
```

Fig 7: Successful creation of the output

```
1.  <% !
2.  String a ="this is how!" ;
3.  %>
4.  <%
5.  for(xx=2;xx<=10;xx=xx+2){
6.  System.out.println(  "the value is : " +
    xx +"");
7.  }
8.  try{
9.  Class.forName("com.mysql.jdbc.Driver
    ");
10. Connection conn=
    DriverManager.getConnection("jdbc:m
    ysql://localhost/demo1","root","passwor
    d123");
11. PreparedStatement ps=
    conn.prepareStatement("Update emp set
    phone=? and sal=? where eid=1011");
12. String b="";
13. b=request.getParameter("t1");
14. ps.setString(1,b);
15. ps.setInt(2,20000);
16. int r=0;
17. r=ps.executeUpdate();
18. if( r!=0) {
19. System.out.println("Update
    successfull");
20. }
21. else {
22. System.out.println("Update
    unsuccessful");
23. }
24. }
25. catch(Exception
    e){e.printStackTrace();}
26. %>
```

Fig 8: Output JSP file

The Fig. 6 shows an input XML file and Fig. 8 shows the output JSP file. The input file has been designed to update certain columns of a database table.

As proposed the input file is checked for well-formedness and validity against the XSD. The xml file is successfully validated as seen in Fig. 7. Once validated the input is then passed on to the operation handler. As previously described the handler performs certain actions specific to the given category of tags.

In the given example, lines 3-5 represent the declaration part of JSP scripting, corresponding to which we get lines 1-3 in the output file. The variable declaration is performed by a variable handler. After the declaration part is done, every other element comes under the scriptlet part of JSP (lines 6-36).
Lines 6-11 represent a standard for-loop ,within which there is a single print statement. The JSP code for it has been generated using a statement and input-output handler respectively.
Lines 12-19 in the input file are used to open a database connection. The data enclosed within these tags is passed on to the database handler, which generated corresponding JSP code(lines 9-10) in Fig. 8.
PreparedStatement is used in JSP to write parameterized sql queries and send different parameters by using same sql queries. This statement is represented by lines 20-31 and its equivalent JSP code is generated (lines 11-15) in Fig. 8.
 In the end of the XML file, to check if the sql query has been successfully executed, an if statement has been used. The equivalent code is then represented by lines 18-22 in Fig. 8. The if statements is processed by the statement handler and in case of improper code a user-defined exception is generated .

For the operation handler checkpoints has been kept to ensure that all the loops, if-else block and other statements are properly closed.

## CONCLUSION & FUTURE WORK

In this paper, we have tried to present a translation process where the user presents the business logic of the pseudo code as an input, and the output is an implementation of the pseudo code in a specific scripting language JSP .The pseudo code is specified in an XML file following the basic rules of XML and conforming to the designed XSD.
This method is simple to understand and can be used for other scripting languages as well. An enhancement to this method can be to add intellisense technology which helps us to reduce typos as well as other common mistakes and to compile the JSP page.